\documentclass[5p,times]{elsarticle}

\usepackage{lineno,bm,amsmath,amssymb,mathtools,booktabs}
\usepackage[mathscr]{eucal}
\modulolinenumbers[1]

\newcommand{\alphadyn}{\alpha_{\rm dyn}}
\newcommand{\avg}[1]{\left\langle#1\right\rangle}
\newcommand{\basic}[1]{\bar{#1}}
\newcommand{\dd}{{\rm d}}
\newcommand{\dr}[1]{\frac{\dd#1}{\dd r}}
\newcommand{\Eq}[1]{\eqref{#1}}

\newcommand{\Fig}[1]{Fig.\,\ref{#1}}
\newcommand{\FF}[1]{Figure\,\ref{#1}}
\newcommand{\glm}{g_l^m}
\newcommand{\hlm}{h_l^m}
\newcommand{\jrms}{j_{\rm rms}}
\newcommand{\Plm}{P_l^m}
\newcommand{\pp}[2]{\frac{\partial#1}{\partial#2}}
\newcommand{\rdyn}{r_{\rm dyn}}
\newcommand{\rlowes}{r_{\rm lowes}}
\newcommand{\rJ}{r_{\rm J}}
\newcommand{\rin}{r_{\rm in}}
\newcommand{\rout}{r_{\rm out}}
\newcommand{\sect}[1]{section~\ref{#1}}

\newcommand{\V}[1]{\boldsymbol{#1}}
\newcommand{\VPhilm}{\V\Phi_l^m}
\newcommand{\VPsilm}{\V\Psi_l^m}
\newcommand{\VYlm}{\V Y_l^m}
\newcommand{\Ylm}{Y_l^m}

\usepackage{amssymb}
\usepackage{color}

\journal{Earth and Planetary Science Letters}

\biboptions{authoryear}

\begin{document}

\begin{frontmatter}

\title{Characterising Jupiter's dynamo radius using its magnetic energy spectrum}

\author[]{Yue-Kin Tsang\corref{cor1}}
\ead{y.tsang@leeds.ac.uk}
\cortext[cor1]{Corresponding author}

\author{Chris A. Jones}

\address{School of Mathematics, University of Leeds, Leeds, LS2 9JT, United Kingdom}

\begin{abstract}
Jupiter's magnetic field is generated by the convection of liquid metallic hydrogen in its interior. The transition from molecular hydrogen to metallic hydrogen as temperature and pressure increase is believed to be a smooth one. As a result, the electrical conductivity in Jupiter varies continuously from being negligible at the surface to a large value in the deeper region. Thus, unlike the Earth where the upper boundary of the dynamo---the dynamo radius---is definitively located at the core-mantle boundary, it is not clear at what depth dynamo action becomes significant in Jupiter. In this paper, using a numerical model of the Jovian dynamo, we examine the magnetic energy spectrum at different depth and identify a dynamo radius below which (and away from the deep inner core) the shape of the magnetic energy spectrum becomes invariant. We find that this shift in the behaviour of the magnetic energy spectrum signifies a change in the dynamics of the system as electric current becomes important. Traditionally, a characteristic radius derived from the Lowes--Mauersberger spectrum---the Lowes radius---gives a good estimate to the Earth's core-mantle boundary. We argue that in our model, the Lowes radius provides a lower bound to the dynamo radius. We also compare the Lowes--Mauersberger spectrum in our model to that obtained from recent Juno observations. The Lowes radius derived from the Juno data is significantly lower than that obtained from our models. The existence of a stably stratified region in the neighbourhood of the transition zone might provide an explanation of this result.
\end{abstract}


\end{frontmatter}

\section{Introduction}
\label{intro}

Jupiter has the strongest magnetic field among the planets in the Solar System. The magnitude of its surface magnetic field is about ten times larger than that of the Earth. Jupiter and the Earth both have dipole-dominated magnetic fields, with the dipolar axis inclined at about 10$^\circ$ to the rotation axis. However, the recent NASA Juno mission \citep{Bolton17} revealed that Jupiter's magnetic field has its non-dipolar part mostly confined to the northern hemisphere \citep{Moore18,Jones18}, unlike the Earth's field which shows no such preference. The intricate magnetic field of Jupiter is believed to be generated by the convective stirring of liquid metallic hydrogen in the planet's interior. An important and long-standing question is at what depth does such dynamo action begin. The dynamo radius---the location of the top of the dynamo region---is an important factor in understanding the interaction between the interior magnetohydrodynamics and the atmospheric flow in the outer layer \citep{Cao17}. Knowledge of the dynamo radius also provides constraints that help to improve the estimation of electrical properties inside Jupiter and will subsequently lead to better modelling of the Jovian magnetic field. The dynamo radius also determines where internal torsional oscillations in Jupiter are reflected as they propagate outwards \citep{Hori19}.

\begin{figure*}
\centering
\includegraphics[width=0.97\textwidth]{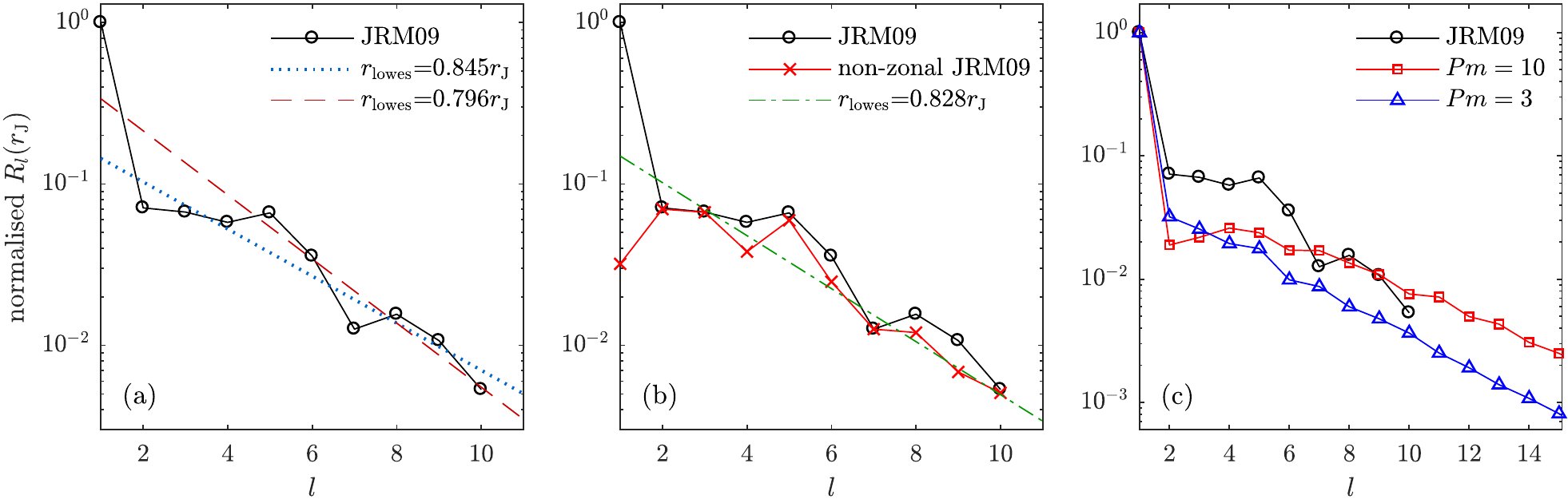}
\caption{(a) Normalised Lowes spectrum $R_l/R_1$ at $r=\rJ$ calculated from the Gauss coefficients in the JRM09 model of \cite{Connerney18}. A linear fit to $\log_{10}R_l(\rJ)$ for $2\leq l \leq10$ gives the Lowes radius $\rlowes=0.845\rJ$. Changing the fitting range to $5\leq l \leq10$ results in $\rlowes=0.796\rJ$. \cite{Connerney18} suggest the data is compatible with $\rlowes=0.87\rJ$. Here, $\rJ=6.9894\times 10^7$m. (b) The non-zonal part of the JRM09 data is compared with the full spectrum. Both spectra are normalised by the value of $R_1$ of the full spectrum. Note that the non-zonal data gives a much closer fit to a straight line in the range $6 \le l \le 10$ than the full data and a linear fit in this range gives $\rlowes=0.828\rJ$. (c) Comparison of normalised Lowes spectrum $R_l/R_1$ at $r=\rJ$ from the Juno data JRM09 and our Jupiter dynamo model at $Pm=10$ and $Pm=3$.}
\label{jrm09}
\end{figure*}

As existing technology does not allow us to take direct measurement inside Jupiter, or for that matter, inside the Earth, we have to deduce the dynamo radius of a planet from measurements made near its surface. In the case of the Earth, where the dynamo radius is at the core-mantle boundary, \cite{Lowes74} introduced a strategy by considering the average magnetic energy over a spherical surface of radius $r$,
\begin{equation}
E_B(r) = \frac{1}{2\mu_0}\frac{1}{4\pi} \oint|\V B(r,\theta,\phi)|^2 \sin\theta\,\dd\theta\,\dd\phi.
\label{eb}
\end{equation}
Here $\V B$ is the magnetic field, $(r,\theta,\phi)$ are the standard spherical coordinates based on the rotation axis, $\mu_0$ is the permeability of free space and the time argument $t$ has been suppressed. From the bottom of the insulating mantle up to the planetary surface, there is no electric current $\V j = \V 0$. The magnetic field in this region can thus be written as $\V B=-\nabla V$. The scalar potential $V$ satisfies $\nabla^2V=0$ and is given by,
\begin{multline}
V(r,\theta,\phi) = a \sum_{l=1}^\infty \sum_{m=0}^l \left(\frac ar \right)^{l+1}\Plm(\cos\theta) \\
\times (\glm \cos m\phi + \hlm\sin m\phi)
\label{vv}
\end{multline}
where $a$ is a reference radius often taken to be the mean planetary radius. $\Plm$ are the Schmidt semi-normalised associated Legendre polynomials. The Gauss coefficients $\glm$ and $\hlm$ are determined from magnetic field measurement at the surface. Using the expression \Eq{vv} in \Eq{eb} yields
\begin{equation}
2\mu_0 E_B(r) = \sum_{l=1}^{\infty} R_l(r)
\label{Rl0}
\end{equation}
where
\begin{equation}
R_l(r) = \left(\frac ar \right)^{2l+4}(l+1) \sum_{m=0}^l\left[ (\glm)^2 + (\hlm)^2 \right]
\label{Rl1}
\end{equation}
is the Lowes spectrum, or sometimes the Lowes--Mauersberger spectrum \citep{Mauersberger56}. It follows that
\begin{equation}
R_l(r) = \left(\frac ar \right)^{2l+4} R_l(a).
\label{dwncont}
\end{equation}
The downward continuation relation \Eq{dwncont} gives the Lowes spectrum $R_l(r)$ at some depth $r$ in terms of $R_l(a)$ at the surface. It relies crucially on $\V B$ being purely potential.

To estimate the depth of the dynamo region, we need one further assumption. It has been argued that the large-scale part of $R_l(a)$ mainly originates from the Earth's outer core and turbulence there results in a uniform distribution of magnetic energy over different scales $l$. In particular, at some depth $\rlowes$ near the core-mantle boundary, $R_l(\rlowes)$ is independent of $l$. This `white source hypothesis' \citep{Backus96}, together with \Eq{dwncont} implies the linear relation 
\begin{equation}
\log_{10} R_l(a)\sim -\beta(a) l
\label{betadef}
\end{equation}
for the large scales with $\beta(a)$ satisfying
\begin{equation}
\rlowes = 10^{-\beta(a)/2} \cdot a.
\label{rlowes}
\end{equation}
Thus, \Eq{rlowes} gives the Lowes radius $\rlowes$ in terms of the spectral slope $\beta$ which can be determined solely from magnetic measurement at the surface. The Lowes radius provides an estimate to the location of the Earth's core-mantle boundary that agrees reasonably with seismic measurement. \cite{Langlais14} found $\rlowes=3294.5$\,km compared to the seismically determined $3481.7$\,km. \cite{ Langlais14} also found that omitting the $m=0$ axisymmetric components in \Eq{Rl1}, so that only the non-zonal components are used,
\begin{equation}
R^{nz}_l(r) = \left(\frac ar \right)^{2l+4}(l+1) \sum_{m=1}^l\left[ (\glm)^2 + (\hlm)^2 \right],
\label{Rl1nonax}
\end{equation}
reduced the scatter of the spectrum and led to a remarkably accurate agreement between $\rlowes$ and the Earth's seismic core radius.

Compared to the Earth, magnetic field measurements for Jupiter are less extensive. The data available before the Juno mission only allowed for the calculation of the Lowes spectrum up to $l=4$ \citep{Connerney98} or $l=7$ \citep{Ridley16} depending on the modelling methodology. This has changed since the Juno spacecraft arrived at Jupiter. While the spacecraft is taking more measurements as it continues to orbit Jupiter, \cite{Connerney18} computed $\glm$ and $\hlm$ up to $l=10$ from the data collected during eight of the first nine flybys. Using these Gauss coefficients, \Fig{jrm09}(a) shows the Lowes spectrum at $r=\rJ$ where, following \cite{French12}, we take
\begin{equation}
\rJ=6.9894\times 10^7\text{m}
\end{equation}
to be the mean radius of Jupiter. Note that \cite{Connerney18} used the equatorial radius, $7.1492 \times 10^7$\,m, and we have corrected for this difference in \Fig{jrm09} and Table~1. In sharp contrast to its Earth counterpart \citep{Backus96}, the Jovian Lowes spectrum $R_l(\rJ)$ does not show a clean exponential decay. In fact $R_l(\rJ)$ remains almost constant for $2\leq l \leq 5$ before decaying at larger $l$. Consequently, routinely applying Lowes' procedure gives different values of $\rlowes$ depending on the range of $l$ used in the linear fit, as shown in \Fig{jrm09}(a). However, using the non-zonal components only on the JRM09 data, see equation \Eq{Rl1nonax}, as suggested by \cite{Langlais14} for the Earth, leads to a better linear fit for $l \ge 6$, see \Fig{jrm09}(b). The best-fit value of $\rlowes=0.828\rJ$. This improvement may arise because the higher order non-axisymmetric field components arise more directly from the non-axisymmetric convection and hence are more randomly distributed than the full spectrum components.

\begin{figure}
\centering
\includegraphics[width=0.41\textwidth]{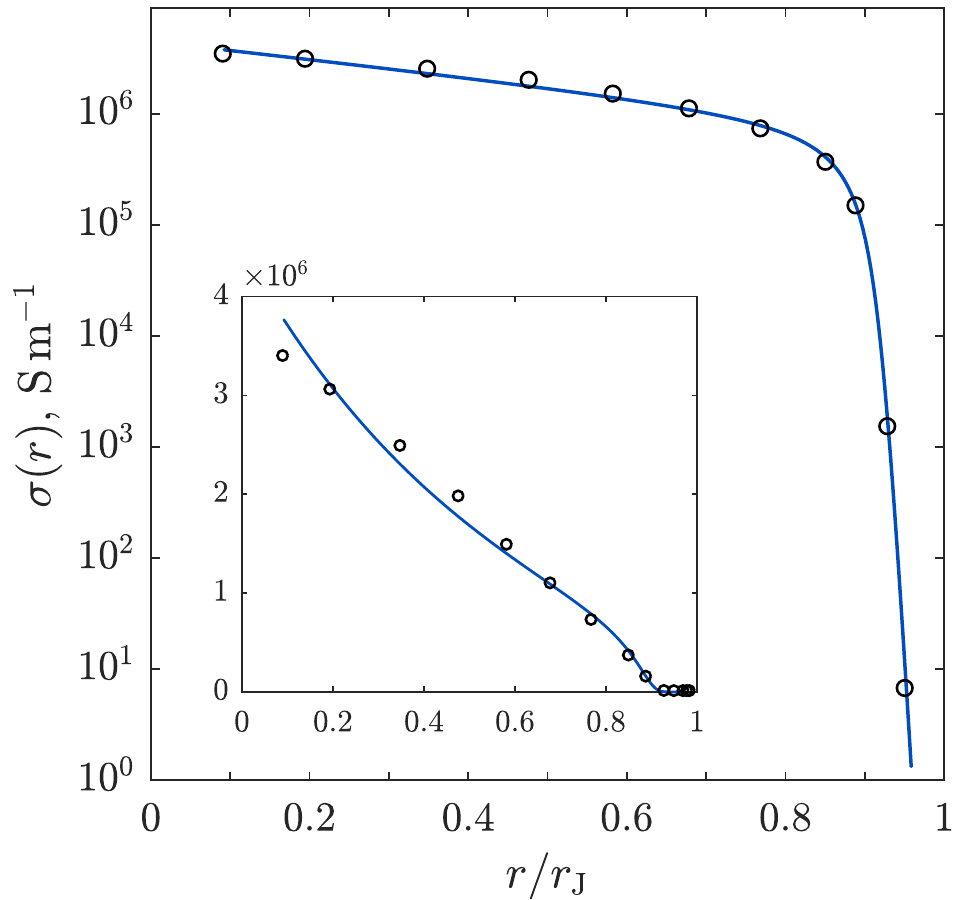}
\caption{Electrical conductivity as a function of depth $r$ along a Jupiter adiabat. The circles are the J11-8a model from the ab initio simulation of \cite{French12}. The solid line is the hyperbolic model \Eq{hyp_sig}. The values of the parameters are $c_1=-4.279\times 10^{-6}$, $c_2=274.9$, $c_3=-2.544\times 10^{-8}$, $c_4=1.801$ and $c_5=20.28$. The inset plots the same data in linear scale.}
\label{sigma}
\end{figure}

\sloppy{
A more fundamental issue here concerns the interpretation of $\rlowes$ for Jupiter. The interior structure of Jupiter is very different from that of the Earth. Theoretical and experimental studies suggest that the phase transition from molecular to  degenerate metallic hydrogen \citep{Wigner35} along a Jupiter adiabat is continuous \citep{Wicht18,Helled18}. As a result, the electrical conductivity $\sigma(r)$ of the hydrogen-helium mixture in Jupiter varies smoothly with the radial distance \citep{Weir96,Liu2008}. For example, \Fig{sigma} shows the profile $\sigma(r)$ obtained from an ab initio simulation by \cite{French12}.} Therefore unlike the Earth, where the flow and the Lorentz force acting on the flow are both confined within the same region, Jupiter's dynamo is coupled to an outer layer of fluid flow that is free from magnetic effects. Such coupling, with a transition layer in between, is not well understood. It is not clear how large the current-free region where the downward continuation operation \Eq{dwncont} is justified actually is. There is also the question about the validity of the white source hypothesis. In fact, how do we characterise the extent of the dynamo region for a continuously varying electrical conductivity profile? Is there a sensible way to define a dynamo radius for Jupiter? In this paper, we examine these issues by considering the magnetic energy spectrum $F_l(r)$, to be defined in \sect{sec:Fl}, in a numerical model of Jupiter. The magnetic energy spectrum $F_l(r)$ essentially represents the distribution of magnetic energy over different spherical harmonic degrees $l$ at depth $r$. The Lowes spectrum $R_l(r)$ in \Eq{Rl1} is a special case of $F_l(r)$ under the condition $\V j = \V 0$ (which is only true near the planetary surface). The change in behaviour of $F_l(r)$ along $r$ indicates varying dynamics in different regions. Comparing $F_l(r)$ to $R_l(r)$ gives further insights into the different physics in these regions.

In the next section, we describe our model for Jupiter's dynamo. In \sect{sec:Fl}, we first introduce the magnetic energy spectrum and discuss how its behaviour changes with depth. We then show that a dynamo radius can be identified from a transition in the spectral slope. In \sect{conclude}, we look at the relationship between the dynamo radius and the Lowes radius. We then finish with a discussion on the differences between results from our model and observation.

\section{A model of Jupiter's dynamo}
\label{model}

Numerical models of Jupiter's dynamo have recently been developed to study the magnetic field and internal flow of the giant planet \citep{Jones11,Jones14,Gastine14}. The model used in the present study is developed and described in detail by \cite{Jones14}, though here the range of parameters has been extended to get further into the strong-field dynamo regime \citep{Dormy16}. We briefly summarise it here.

\subsection{Anelastic spherical dynamo}

We consider the convection of an electrically conducting fluid in a rotating spherical shell of inner radius $\rin=0.092\rJ$ and outer radius $\rout=0.959\rJ$. The heat flux is modelled by an entropy flux proportional to the local entropy gradient with constant diffusivity $\kappa_S$. The other physical parameters are the angular speed $\Omega$, the constant kinematic viscosity $\nu$ and the magnetic diffusivity $\eta(r)$ which varies with the radial distance $r$. The dynamical variables of velocity, magnetic field, entropy, density and pressure $(\V u,\V B, S, \rho, p)$ are governed by the non-dimensional equations,
\begin{subequations}
\begin{gather}
\begin{aligned}
\!\!\! \frac{Ek}{Pm}\frac{{\rm D} \V u}{{\rm D} t} + 2\V{\hat z} \times \V u  = {}
& -\nabla \Pi' - \bigg(\frac{EkRaPm}{Pr}\bigg) S'\dr{\basic T}\V{\hat r} \\
& + \frac{1}{\basic \rho}(\nabla\times\V B)\times\V B + Ek \frac{\V F_{\nu}}{\basic \rho},
\end{aligned}
\label{ueq} \\
\pp{\V B}{t} = \nabla\times(\V u\times \V B) - \nabla\times(\eta\nabla\times \V B), \\[0.1cm]
\begin{aligned}
\basic \rho \basic T \frac{{\rm D} S}{{\rm D} t} = {} & \frac{Pm}{Pr}\nabla\cdot(\basic \rho \basic T \nabla S) \\
& + \frac{Pm}{Pr} \basic \rho \basic T H_S + \frac{Pr}{RaPm} \bigg( Q_\nu + \frac{1}{Ek} Q_J \bigg),
\end{aligned}
\label{heat} \\[0.1cm]
\nabla\cdot(\basic \rho\V u) = 0, \\[0.1cm]
\nabla \cdot \V B = 0,
\end{gather}
\label{aneq}
\end{subequations}
together with the equation of state
\begin{equation}
\dd\rho = \bigg(\frac{\partial\rho}{\partial S}\bigg)_{\!\!p} \dd S + \bigg(\frac{\partial\rho}{\partial p}\bigg)_{\!\!S} \dd p
= \frac{\basic\rho}{\basic g} \dr{\basic T} \dd S - \frac{1}{\basic\rho\basic g} \dr{\basic\rho} \dd p.
\label{eos}
\end{equation}
In deriving the model, the \cite{Lantz99} formulation of the anelastic approximation \citep{Braginsky07} has been employed about a spherically symmetric, hydrostatic and adiabatic basic state $(\basic\rho,\basic p,\basic S)$. This simplifies the system to involve the dynamics of a single thermodynamical variable $S'=S-\basic S$. In \Eq{ueq}, $\V{\hat z}$ and $\V{\hat r}$ are unit vectors along the rotation axis and the radial direction respectively, $\Pi'=p'/\bar\rho+\Phi'$ is a generalised pressure combining disturbance pressure $p'=p-\bar p$ and disturbance gravitational potential $\Phi'$. In \Eq{eos}, $\basic g$ is the gravitational acceleration due to $\basic \rho$. The system is forced by a constant entropy source $H_S$ modelling the secular cooling of the planet. The dissipative terms are:
\begin{subequations}
\begin{gather}
F_{\nu,i} = \sum_{j=1}^3 \pp{}{x_j}\bigg[\basic\rho\bigg(\pp{u_j}{x_i} + \pp{u_i}{x_j}\bigg)\bigg] - \frac23 \pp{}{x_i} (\basic\rho \nabla\cdot\V u), \\
Q_\nu = \bar\rho\bigg[\frac12\sum_{i,j=1}^3\bigg(\pp{u_j}{x_i} + \pp{u_i}{x_j}\bigg)^{\!\!2} - \frac23(\nabla\cdot\V u)^2 \bigg], \\
Q_J = \eta |\nabla\times\V B|^2.
\end{gather}
\end{subequations}
Let $T_*$, $\rho_*$ and $\eta_*$ be the dimensional values of $\basic T$, $\basic \rho$ and $\eta$, respectively, at the midpoint of the shell and $\Delta S$ be the entropy drop across the thickness $L=\rout-\rin$ of the shell. Our equations are non-dimensionalised using the unit of length $L$, time $L^2/\eta_*$ and magnetic field $\sqrt{\Omega \rho_* \mu_0 \eta_*}$, where $\mu_0$ is the permeability of free space. The dimensionless numbers in \Eq{aneq} are defined as:
\begin{align}
Ra = \frac{T_*L^2\Delta S}{\nu\kappa_S},\ Ek = \frac{\nu}{\Omega L^2},\ Pr = \frac{\nu}{\kappa_S},\ Pm = \frac{\nu}{\eta_*}.
\end{align}
The boundary condition for $\V u$ is no-slip at $\rin$ and stress-free at $\rout$. At both the inner and outer boundaries, it is electrically insulating and $S$ is fixed at a constant value. The initial conditions are $\V u=\V 0$ with a small perturbation in $\V B$ and $S$. The spectrum of the initial magnetic perturbation is narrow-banded with $8\leq l\leq 10$ and thus has no dipole component.

\subsection{Hyperbolic electrical conductivity profile}

In applying the anelastic convective system described above to model the Jovian dynamo, we need to provide an equilibrium state and a conductivity profile that represent the thermodynamic and transport properties inside Jupiter. \cite{French12} have calculated the material properties of a hydrogen-helium-water mixture under Jupiter-like condition using density functional theory. Here, we use the same equilibrium density $\basic\rho(r)$ and temperature $\basic T(r)$ profile as in \cite{Jones14} which are smooth interpolations to the J11-8a data in \cite{French12}. For the magnetic diffusivity $\eta(r)$, we consider the following hyperbolic model:
\begin{equation}
(\ln\eta + c_1 r + c_2)(\ln\eta + c_3 r + c_4) = c_5
\label{hyp_sig}
\end{equation}
with five parameters. The values used for these parameters are the same as in \cite{Jones14} which give a good fit to the J11-8a data. \FF{sigma} shows the corresponding electrical conductivity $\sigma(r)=1/\mu_0\eta$. In the interior, $\sigma(r)$ decreases roughly linearly and reaches about one-fifth of its maximum value at $r=0.8\rJ$. This is unlike some of the previous studies \citep{Gastine14,Glatzmaier18} in which the electrical conductivity is taken to be constant below a certain depth. \cite{Dietrich18} studied a wide range of profiles for $\sigma(r)$ by varying $c_2$ in \Eq{hyp_sig} and found a diversity of magnetic field morphologies.

\subsection{Simulation parameters and electric current profile}

For the rest of this paper, the following parameters are kept fixed: $\rin/\rout=0.0963$, $Ra=2\times 10^7$, $Ek=1.5\times 10^{-5}$ and $Pr=0.1$. Jupiter is believed to have a strong-field dynamo, i.e., its magnetic field strongly influences the flow. Thus, we are interested in cases of large magnetic Prandtl number $Pm$ which produce a strong-field dynamo \citep{Dormy16}. Specifically, we investigate the effects of $Pm$ by comparing simulations with $Pm=10$ and $Pm=3$. The case of $Pm=3$ has previously been studied in detail by \cite{Jones14}. Results here are presented in dimensionless units unless otherwise stated.

\begin{figure}
\centering
\includegraphics[width=0.475\textwidth]{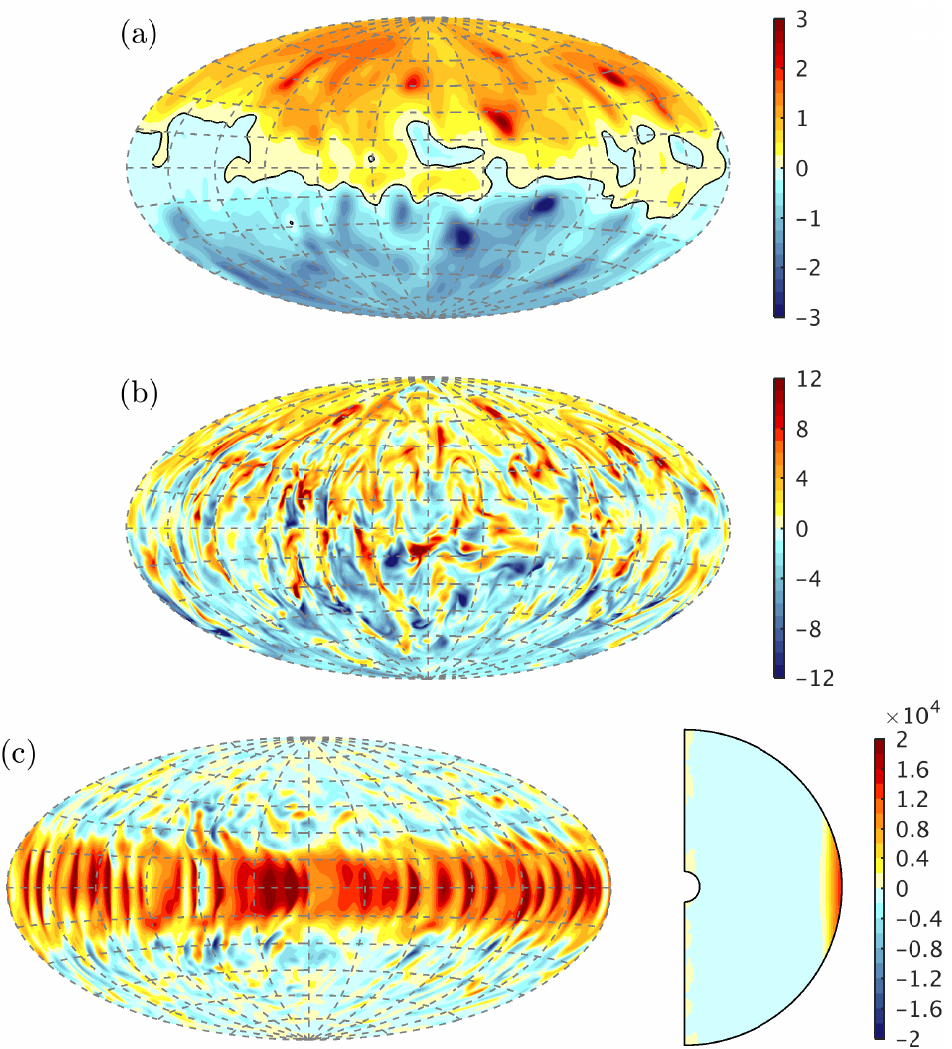}
\caption{Simulation at $Pm=10$. (a) Dipole-dominated radial magnetic field at $r=\rout$. (b) Radial magnetic field at $r=0.8\rJ$ has larger magnitude and more small-scale structures. (c) Left: zonal velocity $u_{\phi}$ at $r=\rout$ showing the prograde equatorial jet. Right: zonal and time averaged $u_{\phi}$.}
\label{ubfields}
\end{figure}

\begin{figure}
\centering
\includegraphics[width=0.475\textwidth]{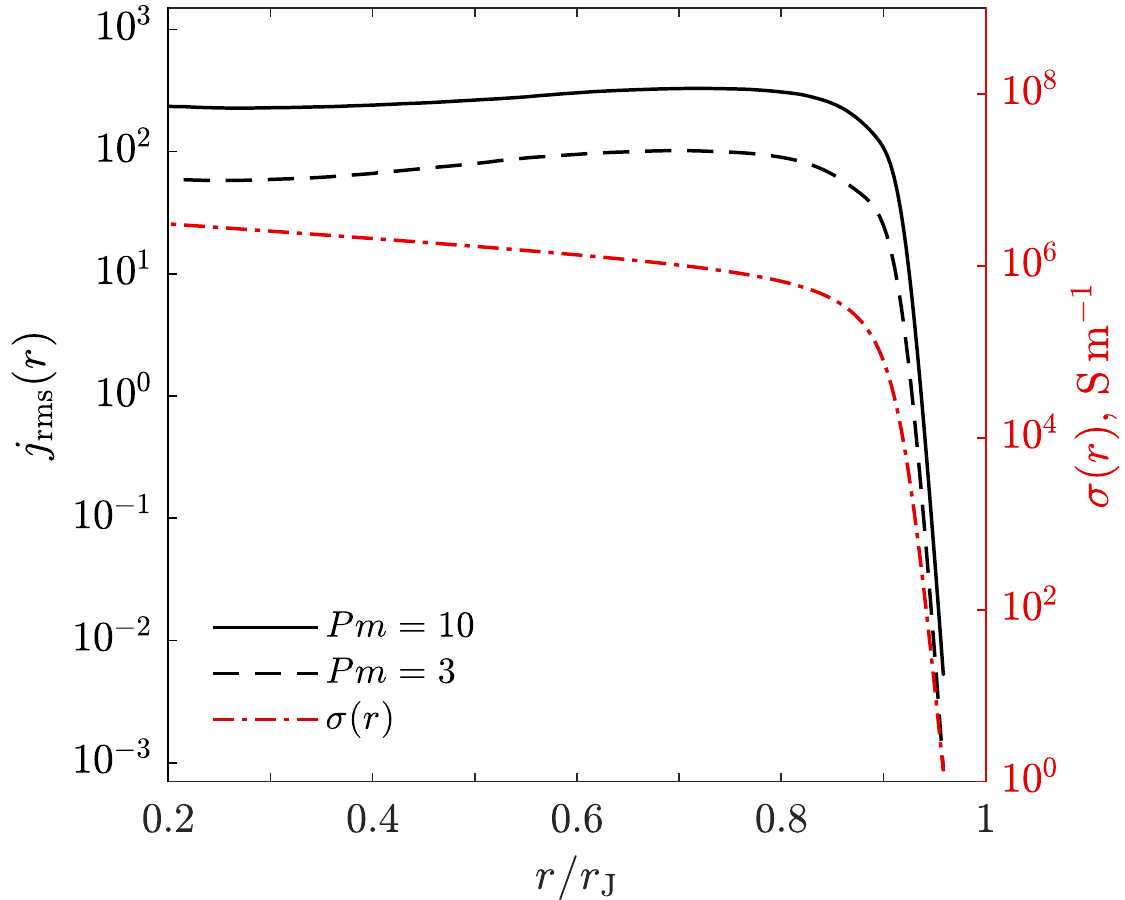}
\caption{Average current $\jrms$, defined in \Eq{jrms}, at different depth $r$ for $Pm=10$ and $Pm=3$. The electrical conductivity $\sigma(r)$ from \Fig{sigma} is also plotted on the right axis.}
\label{current}
\end{figure}

\FF{ubfields} shows snapshots of magnetic and velocity fields from the $Pm=10$ simulation. The equatorial zonal jet and the dipolar nature of the radial magnetic field, both near the surface, are obvious. With the conductivity $\sigma(r)$ increasing sharply with depth, we are interested in the radial dependence of the dynamo action. We first examine the average electric current $j_{\rm rms}$ at a depth $r$, defined as,
\begin{equation}
\jrms^2(r) = \frac{1}{4\pi}\oint \avg{|\V j(r,\theta,\phi,t)|^2}_t \sin\theta\dd\theta\dd\phi.
\label{jrms}
\end{equation}
Above, $\V j=\nabla\times\V B$, $\avg{\cdot}_t$ indicates time average over a statistical steady state and the integral is over a spherical surface of radius $r$. \FF{current} shows that $\jrms(r)$ behaves very similarly for both $Pm=3$ and $Pm=10$. In the interior where strong magnetic field is being generated, $\jrms(r)$ increases slightly with $r$ and peaks at around $r=0.7\rJ$ even though $\sigma(r)$ is monotonically decreasing in $r$. Approaching the surface, $\jrms(r)$ follows the trend of $\sigma(r)$ and drops quickly and smoothly to negligible value, indicating the cessation of dynamo action. While the variation of $\jrms(r)$ with $r$ certainly signifies different dynamics at different depth, \Fig{current} does not locate a characteristic depth that represents the top of the dynamo region. It is not obvious from the profile of $\jrms(r)$  at what depth the electric current becomes large enough to generate a significant magnetic field. In the next section, we show that a dynamo radius can be identified using the magnetic energy spectrum.

\section{Magnetic energy spectrum}
\label{sec:Fl}

We again consider the average magnetic energy on a spherical surface given by \Eq{eb}. Unlike in the derivation of the Lowes spectrum $R_l$ where the magnetic field is assumed to be potential, here we make no such assumption and expand $\V B$ in terms of a set of vector spherical harmonics $\{\VYlm,\VPsilm,\VPhilm\}$ (see \ref{vsh}),
\begin{equation}
\V B = \sum_{l=1}^\infty \sum_{m=-l}^l \big( q_{lm} \VYlm + s_{lm} \VPsilm + t_{lm} \VPhilm \big).
\label{bvsh}
\end{equation}
The expansion coefficients are generally function of $r$ and $t$. Substituting \Eq{bvsh} into \Eq{eb}, we get
\begin{equation}
2 \mu_0 E_B(r,t) = \sum_{l=1}^{\infty} F_l(r,t)
\label{Fldef}
\end{equation}
where the magnetic energy spectrum is
\begin{align}
F_l(r,t) = \frac{4-3\delta_{m,0}}{(2l+1)}\sum_{m=0}^l \big(|q_{lm}|^2 + |s_{lm}|^2 + |t_{lm}|^2\big).
\end{align}
We are mainly interested in the time-averaged spectrum $F_l(r) = \avg{F_l(r,t)}_t$, with the time-averaging done after the system has reached a statistical stationary state. Roughly, $F_l(r)$ can be interpreted as the average magnetic energy per spherical harmonic degree $l$ \citep{Maus08}. Note that $F_l$ is calculated from the full field $\V B$. On the other hand, $R_l$ is calculated from the scalar potential in \Eq{vv}. Nevertheless, in a current-free region, the two are identical. Hence,
\begin{equation}
F_l(r) = R_l(r) \quad \text{if} \quad \jrms(r) = 0.
\end{equation}

\subsection{$F_l(r)$ at the planetary surface}
\label{Flsurf}

We first examine $F_l(r)$ at the outer boundary $r=\rout$ of the spherical shell. \FF{Flt} shows the time evolution of a few modes of $F_l(\rout,t)$ from the $Pm=10$ simulation. The development of the dipolar $l=1$ mode can clearly be seen as it outgrows all other modes.

\begin{figure}
\centering
\includegraphics[width=0.45\textwidth]{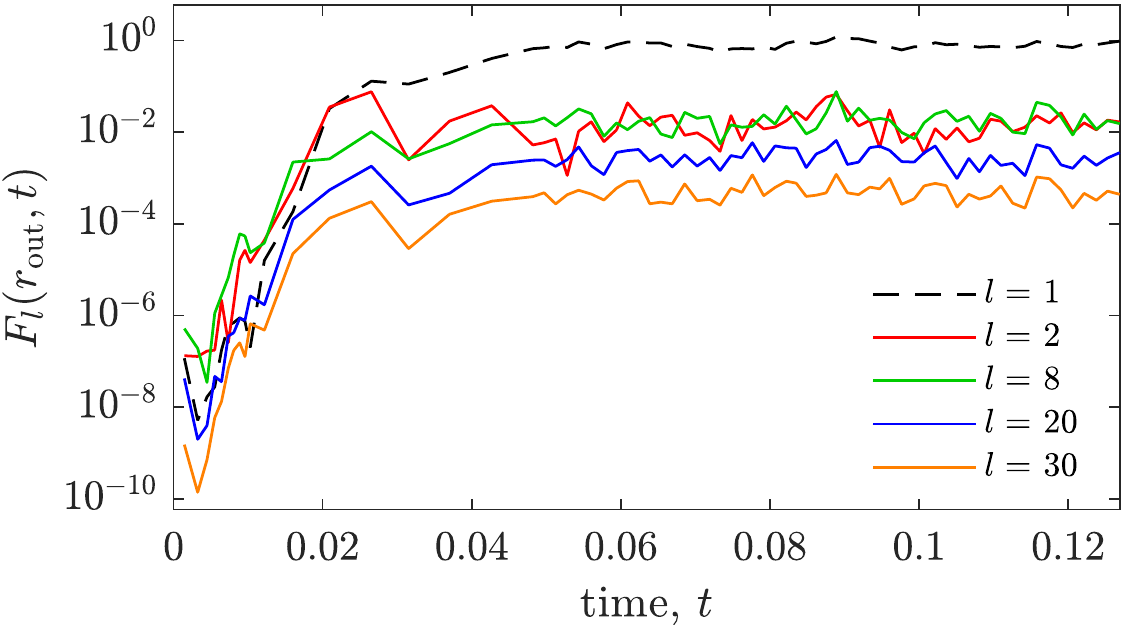}
\caption{Time evolution of selected modes of the magnetic energy spectrum $F_l(\rout,t)$ at the 
outer boundary $r=\rout$ for the case of $Pm=10$. The dipolar $l=1$ mode (dashed line) eventually 
becomes dominant as the system reaches a statistical steady state at about $t=0.05$.}
\label{Flt}
\end{figure}

\begin{table}
\footnotesize
Table~1\\
For simulations at two different $Pm$ and Juno observation (with the two fitting ranges in \Fig{jrm09}(a)): $\alpha(\rout)$ is the spectral slope of $F_l\ (=R_l\text{ at }\rout)$ measured at the outer boundary $\rout$ and $\alphadyn$ is measured inside the 
upper dynamo region as discussed below \Eq{beta_r}. $\rlowes$ is the Lowes radius in \Eq{rlowes} and $\rdyn$ is the dynamo radius in \Eq{rdyndef}. $Rm(r)$ is the depth-dependent magnetic Reynolds number defined in \Eq{Rm} and $r_{Rm}$ is where $Rm=1$. $\rJ=6.9894\times 10^7$m. \\[0.03cm]
\begin{tabular*}{\columnwidth}{@{\extracolsep{\stretch{1}}}lcccccc}
\toprule
                   & $\alpha(\rout)$ & $\alphadyn$ & $\rlowes/\rJ$ & $\rdyn/\rJ$ & $Rm(\rdyn)$ & $r_{Rm}/\rJ$\\
\midrule
$Pm=10$            & 0.072           & 0.024       & 0.883  & 0.907 & 156 & 0.939\\
$Pm=3$             & 0.089           & 0.035       & 0.865  & 0.900 & 111 & 0.934\\
Juno ($l \geq 2$)  & 0.109           & ?           & 0.845  & ? & ? & ?\\
Juno ($l \geq 5$)  & 0.162           & ?           & 0.796  & ? & ? & ?\\
\bottomrule
\end{tabular*}
\end{table}

Since the electric current is negligible at $\rout$, $F_l(\rout) = R_l(\rout)$ in our simulations, see for example \Fig{FlRl} for the case of $Pm=10$. \FF{jrm09}(c) plots $F_l$ from the $Pm=10$ and $Pm=3$ simulations as well as $R_l$  calculated from the Juno data JRM09. All three spectra have been continued to $r=\rJ$ using \Eq{dwncont} and normalised. The key qualitative difference between the spectra from our two simulations is that $F_l(\rJ)$ for $Pm=3$ displays a clear exponential decay for all $l$ (excluding $l=1$) while for $Pm=10$, the spectrum is roughly flat at small $l$ and only starts to decay exponentially for $l\gtrsim 5$. In this respect, the $Pm=10$ spectrum is similar to the Juno spectrum. However, the decay rate of $F_l$ for $Pm=3$ is faster and slightly closer to the Juno observed value. Fitting the range $5\leq l \leq 40$ yields the values of $\rlowes$ shown in Table~1 for $Pm=10$ and $Pm=3$.

\subsection{$F_l(r)$ at different depth $r$}

\begin{figure}
\centering
\includegraphics[width=0.42\textwidth]{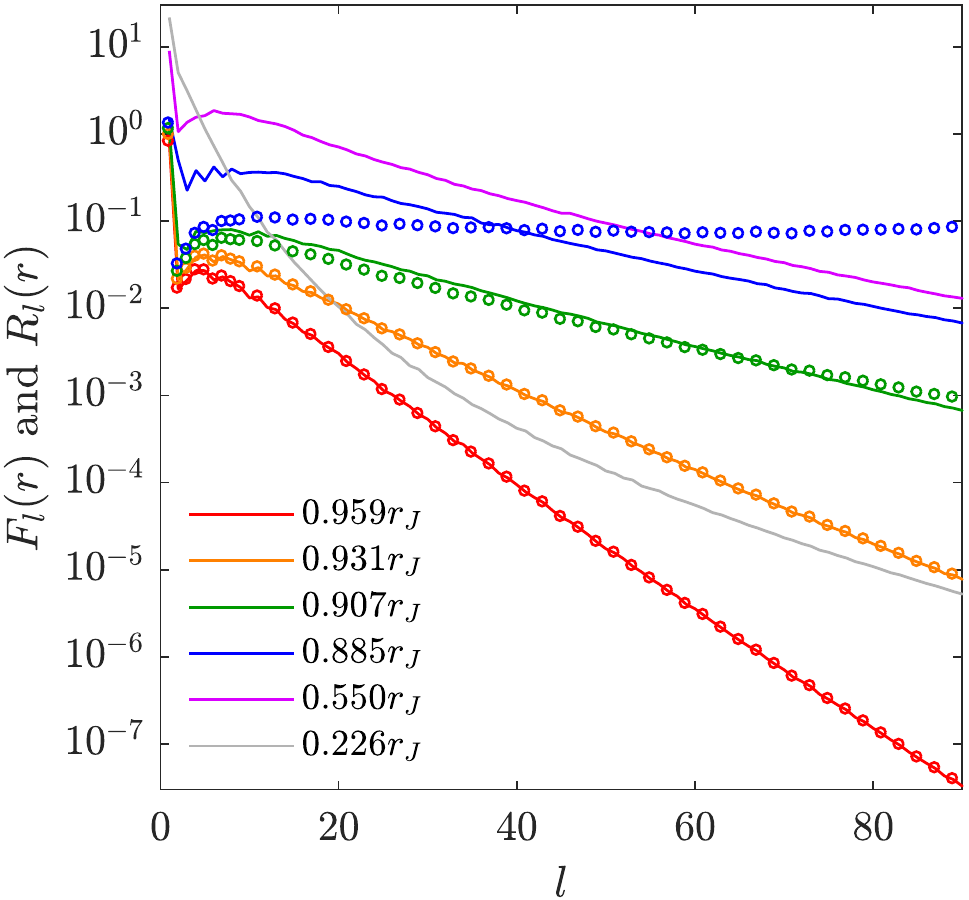}
\caption{Time-averaged magnetic energy spectrum $F_l(r)$ (solid lines), defined in and below \Eq{Fldef}, and Lowes spectrum $R_l(r)$ (circles) at different depth $r$ for the $Pm=10$ simulation. $R_l(r)$ are obtained from the time-averaged $R_l(\rout)$ using \Eq{dwncont}.}
\label{FlRl}
\end{figure}

We now look at the magnetic energy spectrum in the interior of the spherical shell. We focus on the case of $Pm=10$. The solid lines in \Fig{FlRl} show $F_l(r)$ at different depth $r$. The key feature is the transition of $F_l(r)$ through three different stages as $r$ varies. We have seen in the previous section the steep exponential decay of $F_l$ with $l$ (for $l \gtrsim 5$) at the surface. As we move below the surface, \Fig{FlRl} shows $F_l(r)$ maintains such exponential decay,
\begin{equation}
\log_{10} F_l(r) \sim -\alpha(r) l, 
\end{equation}
but the spectral slope $\alpha(r)$ decreases rapidly with $r$ inside the layer of $0.91\rJ \lesssim r \leq \rout$ and $F_l(r)$ has become rather shallow at $0.91\rJ$. As we delve further into the interior, quite remarkably, $F_l(r)$ remains shallow as its shape, and hence $\alpha(r)$,  becomes more or less invariant over the substantial region of $0.55\rJ \lesssim r \lesssim 0.91\rJ$. This clearly indicates a shift in the dynamics near $0.91\rJ$. Finally, in the deep interior and close to the core, $F_l(r)$ decays super-exponentially and the magnetic field is dominated by large scales. Boussinesq geodynamo models which compute the magnetic energy spectrum inside the core also show a spectrum that decays exponentially with $l$ for $l \ge 5$, e.g. \cite{Christensen99}.

\subsection{A dynamo radius}
\label{sec:rdyn}

The significance of the change in behaviour of $F_l(r)$ becomes clear when we compare $F_l(r)$ to the Lowes spectrum $R_l(r)$ at the same depth. Recall that $R_l(\rout)=F_l(\rout)$ at the surface. We now downward continue $R_l(\rout)$ using \Eq{dwncont} (with $a=\rout$) to obtain $R_l(r)$ at different $r$, which we plot as circles in \Fig{FlRl}. Near the surface, $F_l(r)$ and $R_l(r)$ are essentially indistinguishable because electric current is negligible there. As $r$ decreases and reaches some depth $\rdyn$, $F_l(r)$ starts to deviate from $R_l(r)$. The main observation in \Fig{FlRl} is that the shape of $F_l(r)$ becomes independent of $r$, as discussed in the previous section, at essentially the same depth $\rdyn$. This implies that $\rdyn$ is the boundary below which electric current becomes important and the dynamics of the system is altered. We therefore identify $\rdyn$ as the top of the dynamo region, or the dynamo radius.

\begin{figure}
\centering
\includegraphics[width=0.45\textwidth]{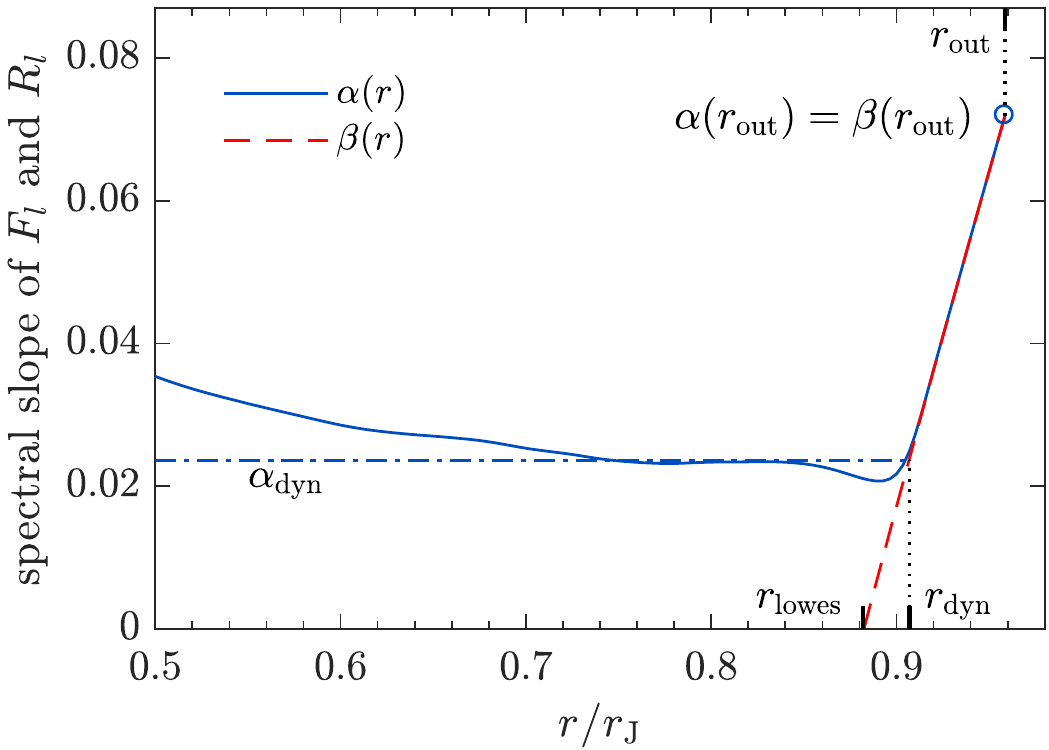}
\caption{Spectral slope $\alpha(r)$ of the magnetic energy spectrum $F_l(r)$ and the spectral slope $\beta(r)$ of the Lowes spectrum $R_l(r)$ in the $Pm=10$ simulation. At the dynamo radius $\rdyn$, $\alpha$ diverges from $\beta$ and remains more or less constant about $\alphadyn$ inside a large part of the dynamo region. The Lowes radius $\rlowes$ is where the downward continued $R_l$ is flat, $\beta(\rlowes)=0$.}
\label{slopeFR}
\end{figure}

\FF{slopeFR} vividly illustrates the discussion in the previous paragraph by plotting $\alpha(r)$ together with the Lowes spectral slope $\beta(r)$, defined analogously to \Eq{betadef}, as a function of $r$. From \Eq{dwncont}, we have
\begin{equation}
\beta(r) = \beta(\rout) - 2 \log_{10}\frac{\rout}{r},
\label{beta_r}
\end{equation}
where $\beta(\rout)=\alpha(\rout)$. Note how $\alpha(r)$ diverges from $\beta(r)$ and levels off to the value $\alphadyn$ at $r=\rdyn$. The sharpness of the transition allows for a meaningful definition of $\rdyn$. \FF{slopeFR} also provides a quantitative way to determine $\rdyn$. Fitting a horizontal line through $0.7\rJ < r < 0.86\rJ$ yields the value of $\alphadyn=0.024$. We then obtain the dynamo radius $\rdyn$ from \Eq{beta_r} using the relation
\begin{equation}
\beta(\rdyn) = \alphadyn.
\label{rdyndef}
\end{equation}
This gives $\rdyn=0.907\rJ$ for the $Pm=10$ simulation. The values of the various spectral slopes and characteristic radii for $Pm=10$ are summarised in Table~1.

Spacecraft missions can only measure the spectral slope at the planetary surface, from which $\rlowes$ is calculated using the white source assumption discussed in \sect{intro}. The question is then how well can $\rlowes$ predict the actual dynamo radius $\rdyn$. In our simulations, $\rlowes$ is where the dashed line in \Fig{slopeFR} intersects the horizontal axis, i.e. $\beta(\rlowes)=0$ in \Eq{beta_r}. This is because by definition, the downward continued $R_l(r)$ becomes flat at $r=\rlowes$, at least for the range of $l$ where $R_l(\rout)$ is fitted to obtain $\beta(\rout)$. It is clear from \Fig{slopeFR} that generally
\begin{equation}
\rlowes \leq \rdyn.
\end{equation}
Comparing the values shown in Table~1, we see that for $Pm=10$, $\rlowes$ is  about 3\% less than $\rdyn$. The difference stems from $\alphadyn$ being fairly small but not exactly zero. Comparing $F_l$ to $R_l$ near $\rlowes$ in \Fig{FlRl} again shows the white source assumption is only approximate and $F_l(r)$ never becomes exactly flat. Nonetheless, helped by the steep decrease of $\beta(r)$ with $r$ shown in \Fig{slopeFR}, we still have a close agreement between $\rlowes$ and $\rdyn$. \FF{ubfields}(b) shows a snapshot of the radial magnetic field at $r=0.8\rJ$ where $\alpha(0.8\rJ)\approx\alphadyn$.
\begin{figure}
\centering
\includegraphics[width=0.47\textwidth]{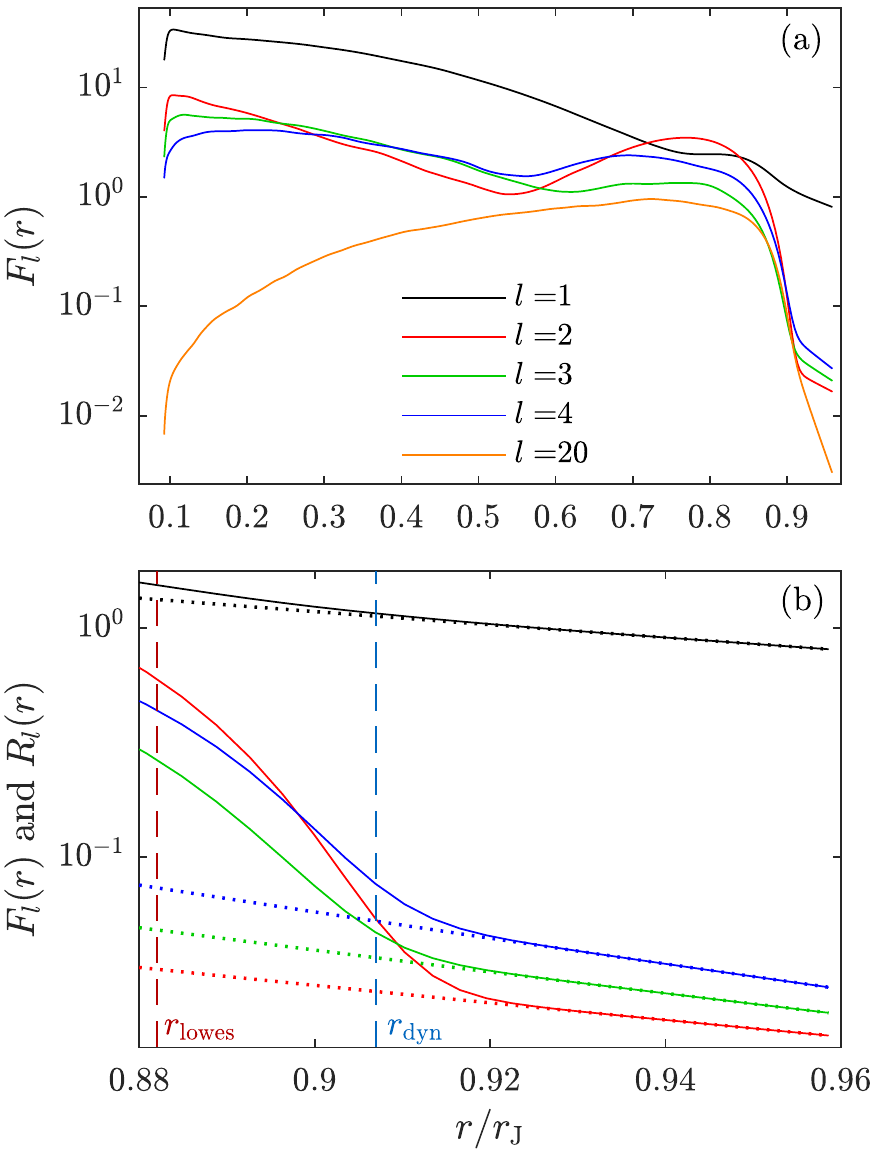}
\caption{Selected modes of the magnetic energy spectrum $F_l(r)$ and the Lowes spectrum $R_l(r)$ as a function of depth $r$ in the simulation at $Pm=10$. (a) $F_l(r)$ over the whole spherical shell. (b) $F_l(r)$ (solid lines) and $R_l(r)$ (dotted lines) near a region about the dynamo radius $\rdyn$.}
\label{Flr}
\end{figure}

\FF{Flr} reveals further details about the radial dependence of the magnetic energy spectrum by plotting a selected number of modes of $F_l(r)$ as a function of $r$. The first few modes ($l \lesssim 5$) are exceptional as they varies irregularly with $r$ while the rest of the modes are well represented by $l=20$ in \Fig{Flr}(a). The $l=1$ mode has the largest magnitude at all $r$ except for a region around $0.8\rJ$ where it is overtaken by $l=2$, see also \Fig{ubfields}(b). Interestingly, this is also the region in which $\jrms(r)$ peaks and $\alpha(r)$ remains virtually constant at $\alphadyn$. We also see that all non-dipolar modes are strongly damped just above the dynamo radius $\rdyn$ leading to the dipole-dominated field observed at the surface. \FF{Flr}(b) zooms into a region about $r=\rdyn$ and shows, for the large-scale modes, how $F_l(r)$ deviates from $R_l(r)$ as one moves from the current-free outer layer into the dynamo region.

\subsection{Effects of $Pm$}  

We now compare the simulation at $Pm=10$ to the one at $Pm=3$. We have already discussed the differences in the magnetic energy spectrum at the surface in \sect{Flsurf}. The slightly steeper $F_l(\rout)$ for $Pm=3$ means the magnetic field has less small scales. Generally, results from the two simulations are qualitatively similar. The $Pm=10$ simulation has stronger magnetic fields, so the Elsasser number, which is proportional to the root-mean-squared magnetic field over the domain, is about an order of magnitude larger than that of the $Pm=3$ simulation, closer to the high value expected in Jupiter. In the $Pm=3$ simulation, it was noted in \cite{Jones14} that the Lorentz force in the interior mostly suppresses the internal differential rotation in the metallic hydrogen region. At $Pm=10$ the stronger magnetic field means the internal differential rotation is even more strongly suppressed, see \Fig{ubfields}(c), so that strong zonal flow occurs only in the molecular region near the equator. However, although the zonal flow is confined to the molecular region, its magnitude is about double that of the $Pm=3$ simulation. The convective velocity in the $Pm=10$ simulation is also about double that in the $Pm=3$ simulation, so the magnetic Reynolds number is about twice that of the $Pm=3$ simulation. While it is not computationally possible to achieve the parameters believed to operate in Jupiter, the increase in $Pm$ has moved the simulation results in the direction of more realistic values.

\FF{current} shows the electric current is roughly three times smaller in the $Pm=3$ case. The spectral slopes for $Pm=3$ display the same trend as in \Fig{slopeFR}. The values of $\alpha(\rout)$ and $\alphadyn$ together with that of $\rlowes$ and $\rdyn$ are given in Table~1. A steeper spectrum at the surface means a bigger $\alpha(\rout)$. At the same time, we see that $\alphadyn$ also increases. The net result is that the dynamo radius $\rdyn$ is only marginally less than that of $Pm=10$. On the other hand, $\rlowes$ drops more significantly which makes $\rlowes$ less accurate as a predictor of $\rdyn$. These results suggest that for a dynamo with a larger $Pm$, the magnetic energy spectrum $F_l(r)$ in the upper part of the dynamo region is closer to being `white'. As a consequence, the Lowes radius gives a better prediction to the dynamo radius.

\section{Discussion and conclusion}
\label{conclude}

The electrical conductivity in Jupiter varies from being negligible at the surface to a very high value in the interior. It thus raises the question about the depth at which dynamo action starts. In this paper, we consider the magnetic energy spectrum $F_l(r)$ at depth $r$ in a numerical model of Jupiter's dynamo. For $l \gtrsim 5$, the magnetic energy spectrum decays exponentially with $l$, $\log_{10} F_l(r) \sim -\alpha(r) l$. We find that a sharp transition in $\alpha(r)$ can be used to identify a dynamo radius $\rdyn$ and this dynamo radius can be reasonably predicted by the Lowes radius $\rlowes$ as discussed in \sect{sec:rdyn}. The situation is in fact rather simple as illustrated in \Fig{slopeFR}. The two characteristic radii $\rdyn$ and $\rlowes$ are controlled by two spectral slopes: $\alpha(\rout)$ which is observable at the surface and $\alphadyn$ which measures the deviation from the white source hypothesis near the top of the dynamo region. Varying $\alphadyn$ moves the horizontal dot-dashed line in \Fig{slopeFR} up and down while changing $\alpha(\rout)$ shifts the dashed curve left and right. These determine the location of $\rdyn$ and $\rlowes$ as well as the relative distance between them. Notice that the dashed curve, given by \Eq{beta_r}, is essentially a straight line for $\rout-r \ll \rout$,
\begin{equation}
\beta(r) \approx \frac{2r}{\rout} -2 + \alpha(\rout).
\end{equation}
We find that in our two simulations at $Pm=10$ and $Pm=3$, $\alpha(\rout)$ and $\alphadyn$ change in such a way that leaves the dynamo radius fairly insensitive to $Pm$. Incidentally, at $r=\rdyn$, the electrical conductivity $\sigma(r)$ has dropped by two orders of magnitude from its maximum at the inner boundary.

\begin{figure}
\centering
\includegraphics[width=0.449\textwidth]{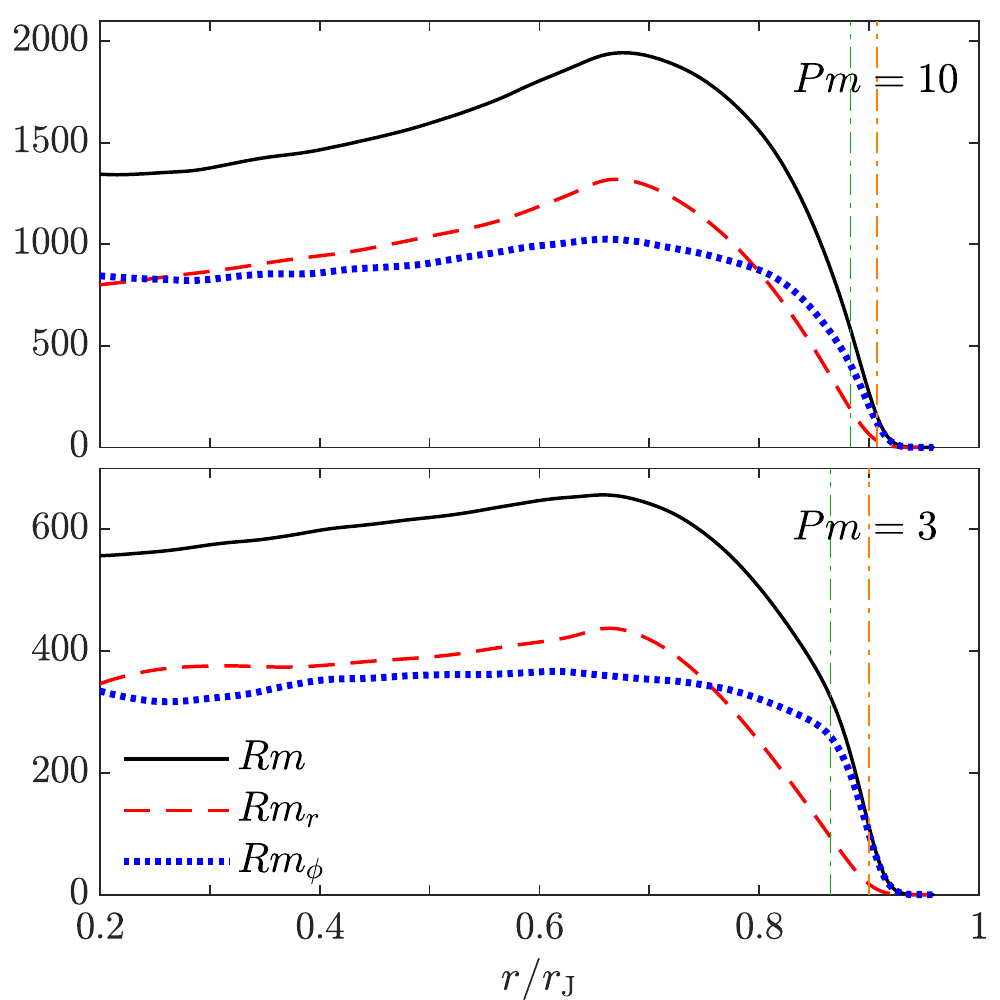}
\caption{Depth-dependent magnetic Reynolds numbers, defined in \Eq{Rm}, based on different velocity scales for $Pm=10$ and $Pm=3$. The two vertical dot-dashed lines indicate the locations of $\rdyn$ and $\rlowes$ given in Table~1.}
\label{Rm_r}
\end{figure}

\FF{Rm_r} plots as a function of depth three different magnetic Reynolds numbers, each based on a different velocity scale,
\begin{equation}
Rm = \frac{L u_{\rm rms}(r)}{\eta(r)}, \quad Rm_r = \frac{L u_{r,\rm rms}(r)}{\eta(r)}, \quad Rm_\phi = \frac{L u_{\phi,\rm rms}(r)}{\eta(r)},
\label{Rm}
\end{equation}
for our two Jovian dynamo simulations, reverting to dimensional units. Here, $u_{\rm rms}$, $u_{r,\rm rms}$ and $u_{\phi,\rm rms}$ are the root-mean-squared values of the total, radial and zonal velocity, respectively, over a spherical surface of radius $r$. All three magnetic Reynolds numbers vary weakly in the interior and then decrease sharply to a negligible value near the surface. We also see that $Rm \approx Rm_\phi$ for $r \gtrsim \rdyn$ indicating the zonal flow becomes dominant. This is consistent with the depth of the equatorial zonal jet estimated roughly from figures such as \Fig{ubfields}. The values of $Rm$ at $\rdyn$ and the depth $r_{Rm}$ at which $Rm=1$ are given in Table~1. These values suggest that $r_{Rm}$ determined by the criterion $Rm=1$ is generally larger than $\rdyn$ estimated from $F_l(r)$. Using $Rm_\phi$ instead of $Rm$ gives similar estimates. However, $Rm_r$ does give a somewhat smaller $r_{Rm}\approx 0.930\rJ$ and $Rm_r(\rdyn)\sim O(10)$ for $Pm=10$. The three magnetic Reynolds numbers in \Eq{Rm} all use the shell thickness $L$ as the typical length scale. An alternative is to use the magnetic diffusivity scale height $d_\eta(r)=\eta(\dd\eta/\dd r)^{-1}$. \cite{Dietrich18} and \cite{Wicht19} have examined, among other things, the characterisation of Jovian dynamo models using different magnetic Reynolds numbers. Since $d_\eta \ll L$, we expect $Rm\sim Rm_\phi \sim O(10)$ and $Rm_r\sim O(1)$ at $\rdyn$ in our simulations if the definition of $Rm$ based on $d_\eta(r)$ is used. On the other hand, numerical dynamo models cannot achieve the large value of $Rm$ expected in the interior of Jupiter. If we adopt the velocity of $U=10^{-2}$\,m\,s$^{-1}$ which has been estimated for Jupiter's interior, see e.g. \citet{Jones14}, and use our values of $L$ and $\eta$, then the value of $r$ at which $Rm=U L / \eta = 1$ increases to $r=0.957r_J$. 

We have estimated the Lowes radius from the Juno data in \Fig{jrm09}. We find that it has fairly large uncertainties depending on the range of $l$ used in the linear fitting. However, if the zonal components are omitted we find a closer linear fit for $l \ge 6$ as did \cite{Langlais14} for the geomagnetic data. The situation in Jupiter could be similar to our simulation at $Pm=10$ where a clean exponential decay in the spectrum emerges only at larger $l$. This is in contrast to the case of $Pm=3$, which has weaker flow and magnetic field. We anticipate further data collected by the ongoing and future flybys will extend the range of the Lowes spectrum in \Fig{jrm09} and hence provide a more reliable estimate of $\rlowes$.

Despite uncertainties in the data, Table~1 shows that $\rlowes$ from the Juno observation is clearly smaller than in both of our simulations. Nevertheless, its implication on the location of the dynamo radius is not clear. Whether $\rlowes$ gives a good estimate on $\rdyn$ relies on the white source hypothesis, which may or may not be valid in Jupiter. However, the $Pm=10$ simulation, which we believe is closer to Jupiter conditions than the $Pm=3$ simulation, has a smaller $\alphadyn$ than the $Pm=3$ simulation, suggesting that Jupiter's magnetic field might be close to white near the top of the dynamo region. In our simulations, a steeper spectrum observed at the surface tends to be accompanied by steeper spectra in the interior and consequently $\rdyn$ could be shallower than $\rlowes$ by a fair amount. The present results suggest that the true dynamo radius likely lies above the Lowes radius.

Irrespective of its relation to the dynamo radius, the Lowes radius is a property of the magnetic field at the surface of Jupiter. The smaller $\rlowes$ of the Juno data stems from a steeper spectrum at the planetary surface, implying Jupiter's magnetic field has less small scales than that in our model. This is slightly surprising as the flow is believed to be more vigorous in Jupiter than in our simulations, because the simulations have enhanced diffusion coefficients to maintain numerical stability. We should point out again that $\rlowes$ for the simulations in Table~1 are derived from time-averaged spectra while the Juno observation essentially provides only a snapshot of Jupiter's magnetic field. \FF{slope_t} plots the instantaneous Lowes radius $\rlowes(t)$ obtained from the time-dependent spectrum $F_l(\rout,t)$ in the $Pm=10$ simulation. The case of $Pm=3$ shows similar spread about the mean value. We argue that the difference between simulations and observation is significant even when statistical fluctuation is taken into account.

\begin{figure}
\centering
\includegraphics[width=0.45\textwidth]{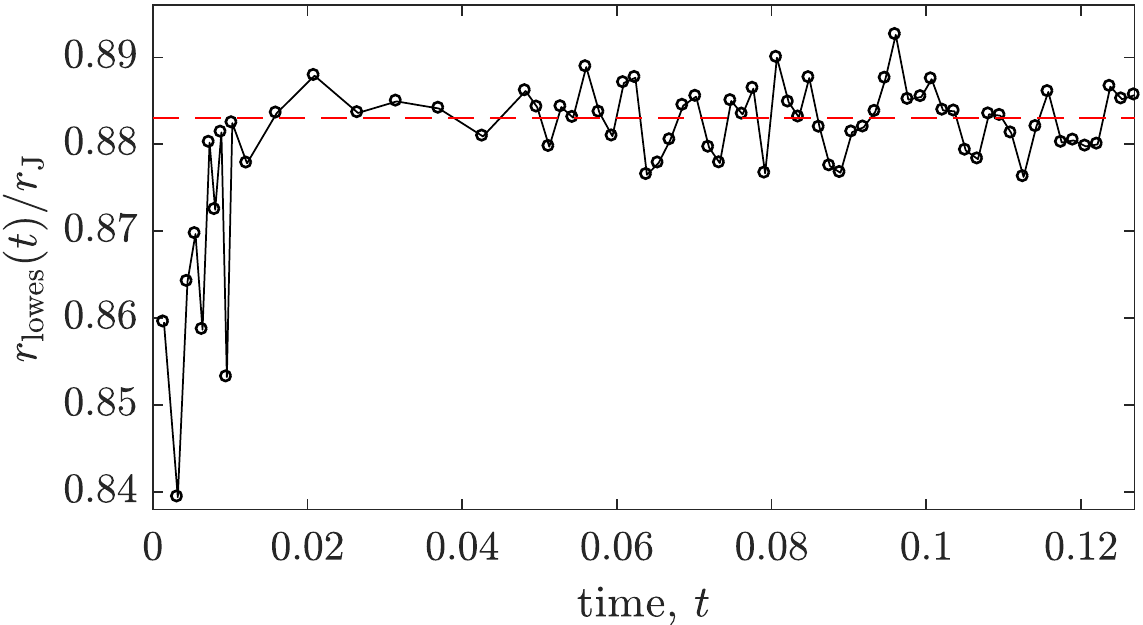}
\caption{Instantaneous Lowes radius $\rlowes(t)$ obtained from the time-dependent magnetic energy spectrum $F_l(\rout,t)$ at the outer boundary for $Pm=10$. The dashed line is the time-averaged value over the statistical steady state given in Table~1.}
\label{slope_t}
\end{figure}

The difference in $\rlowes$ between our numerical model and observation raises several questions and suggests possible avenues for future research. The results presented here are specific to the electrical conductivity profile $\sigma(r)$ in \Eq{hyp_sig} which is based on data from theoretical ab initio calculation. Could the deeper Lowes radius in observation mean the actual electrical conductivity inside Jupiter is smaller than predicted? It is worth studying how the magnetic energy spectrum responds to perturbations in $\sigma(r)$. On a related note, the magnetic Reynolds number $Rm$ in our simulations is of the order $10^3$ at its maximum, much lower than an estimated value of $10^6$ in Jupiter \citep{Jones14}. The puzzle here is that increasing $Rm$ in the model will likely increase $\rlowes$ rather than reduce it and thus move it further away from the Juno value. While current computing resources prevent us from reaching a much larger $Rm$, investigating the trend of the dynamics in the neighbourhood of a smaller attainable $Rm$, possibly by changing $\sigma(r)$, could provide valuable insights.

In our simulations, the system is forced by the constant entropy source $H_S$ in \Eq{heat} and we employ a constant entropy boundary condition. Could our numerical setup tend to produce extra small scales that lead to the shallower spectra shown in \Fig{jrm09}(c)? In geodynamo simulations, boundary conditions can significantly affect the dynamics \citep{Sakuraba09,Dharmaraj12}. It is important to assess the robustness of the present results and examine their dependence on boundary conditions and the form of forcing.

The formation of a stably stratified layer just under the molecular layer due to `helium rain' \citep{Stevenson77} has been proposed to explain the near-axisymmetric magnetic field of Saturn \citep{Stevenson82,Dougherty18}. Although helium rain is more probable to occur in Saturn, it cannot be ruled out for Jupiter \citep{Wahl17,Debras19}. It would be interesting to see the effects of such a stable layer on the Lowes radius as it displaces the dynamo action deeper into the interior. This is perhaps the most natural way to explain the surprisingly low value of $\rlowes$ in the Juno data.

\section*{Acknowledgements}

\sloppy{The authors are supported by the Science and Technology Facilities Council (STFC), `A Consolidated Grant in Astrophysical Fluids' (grant numbers ST/K000853/1 and ST/S00047X/1). This work used the DiRAC@Durham facility managed by the Institute for Computational Cosmology on behalf of the STFC DiRAC HPC Facility (www.dirac.ac.uk). The equipment was funded by BEIS capital funding via STFC capital grants ST/P002293/1, ST/R002371/1 and ST/S002502/1, Durham University and STFC operations grant ST/R000832/1. DiRAC is part of the National e-Infrastructure.}

\appendix

\section{Vector spherical harmonics}
\label{vsh}

Following \cite{Barrera85} but using the Schmidt semi-normalised associated Legendre polynomials $P_l^m$ in the definition of the spherical harmonics,
\begin{equation}
\Ylm(\theta,\phi) = P_l^{|m|}(\cos\theta) e^{im\phi},
\end{equation}
we define three vector spherical harmonics:
\begin{subequations}
\begin{align}
\VYlm(\theta,\phi) &= \Ylm \V{\hat r}, \\
\VPsilm(\theta,\phi) &= \frac{1}{\sqrt{l(l+1)}}\, r \nabla \Ylm, \\
\VPhilm(\theta,\phi) &= \V{\hat r} \times \VPsilm,
\end{align}
\end{subequations}
which form an orthogonal basis for all square-integrable vector fields on the unit sphere. The (semi-)normalisation condition is
\begin{gather}
\oint \VYlm \cdot (\V Y_{l'}^{m'})^* \sin\theta\dd\theta\dd\phi
= \frac{4\pi}{2l+1}(2-\delta_{m,0})\delta_{ll'}\delta_{mm'},
\end{gather}
with similar expressions for $\VPsilm$ and $\VPhilm$.

\bibliography{jupspec}

\end{document}